\documentclass[12pt]{article}
\usepackage{graphicx}
\usepackage{amsmath}
\usepackage{amsfonts}
\usepackage{bm}
\usepackage{slashbox}
\usepackage{hyperref}
\numberwithin{equation}{section}

\def\a{\alpha}
\def\b{\beta}

\def\g{\gamma}

\def\t{\tau}

\def\beq{\begin{eqnarray}}
\def\eeq{\end{eqnarray}}

\newcommand{\vev}[1]{ \left\langle {#1} \right\rangle }

\def\SO{\mathrm{SO}}
\def\SU{\mathrm{SU}}
\def\U{\mathrm{U}}

\def\cN{\mathcal{N}}

\begin{document}

\begin{titlepage}

\begin{flushright}
UT-11-15\\
IPMU-11-0086\\
\end{flushright}

\vskip 1.35cm
\begin{center}

\def\thefootnote{\fnsymbol{footnote}}

{\large \bfseries
$\cN=1$ curves for trifundamentals
}
\vskip 1.2cm
Yuji Tachikawa$^{0}$\footnote[1]{on leave from IPMU}  and Kazuya Yonekura$^{1,2}$
\vskip 0.4cm

{

$^0$ School of Natural Sciences, Institute for Advanced Study, \\
Princeton, New Jersey 08504, USA 

\medskip

$^1$ Institute for the Physics and Mathematics of 
the Universe (IPMU),\\  
University of Tokyo, Chiba 277-8568, Japan\\ 
$^2$  Department of Physics, University of Tokyo,\\
    Tokyo 113-0033, Japan}

\vskip 1.5cm

\abstract{
We study the Coulomb phase of $\cN=1$ $\SU(2)^3$ gauge theory coupled to one trifundamental field, and generalizations thereof. 
The moduli space of vacua is always one-dimensional with multiple unbroken $\U(1)$ fields.
We find that the $\cN=1$ Seiberg-Witten curve which encodes the $\U(1)$ couplings is given by the double cover of a Riemann surface branched at the poles and the zeros of a meromorphic function.
}
\end{center}
\end{titlepage}

\setcounter{tocdepth}{2}
\tableofcontents

\section{Introduction}
The protagonist of this paper is the \emph{trifundamentals}, i.e.~chiral superfields of the form $Q_{\alpha_1\alpha_2\alpha_3}$ ($\alpha_{1,2,3}=1,2$) transforming under $\SU(2)_1\times\SU(2)_2\times\SU(2)_3$ as $(2,2,2)$. 
A trifundamental counts as two flavors of quark superfields from the point of view of one $\SU(2)$, and thus can be used to construct many types of renormalizable supersymmetric gauge theories.
Theories with trifundamentals were studied in the last century e.g.~in \cite{Intriligator:1994sm,Intriligator:1995id,Argyres:1999fc}, but only implicitly and very rarely. 
Presumably, this is because a naive generalization of a trifundamental of $\SU(2)^3$ to $\SU(N)^3$ cannot be used in a renormalizable supersymmetric gauge theory.
The situation changed drastically with the publication of the paper \cite{Gaiotto:2009we} where trifundamentals are finally put to good use: the fact that three $\SU(2)$ symmetries transform the trifundamental in the same way was crucial in the unified treatment of the S-duality of $\cN=2$ gauge theory with multiple $\SU(2)$ gauge groups. 
The similar role for $\SU(N)$ gauge theories is played by a superconformal theory now known as $T_N$ theory which has $\SU(N)^3$ flavor symmetry and does not have a simple Lagrangian description if $N>2$. 
Thus the trifundamental of $\SU(2)$ has found a rightful place as the first and the only free member $T_2$ of the family of $T_N$ theories.

Therefore, it might not be completely useless to explore the dynamics of $\cN=1$ gauge theories with chiral matter fields in the trifundamental representation of $\SU(2)$.   Conformal points of such theories were studied in \cite{Maruyoshi:2009uk,Benini:2009mz}, and our objective is to study the Abelian Coulomb phase of these theories where the low energy limit contains only unbroken $\U(1)$ gauge fields and neutral moduli fields.  
Supersymmetry requires that the physical coupling matrix of the $\U(1)$ fields of $\cN=1$ theory depends holomorphically on the moduli fields. Furthermore, moving along a closed loop in the moduli space can induce electromagnetic duality transformation on the $\U(1)$ fields. 
Then, as in $\cN=2$ case \cite{Seiberg:1994rs,Seiberg:1994aj}, it is convenient to package the data into the complex structure of a Riemann surface, which we call $\cN=1$ curve \cite{Intriligator:1994sm}.\footnote{An almost complete list of references on $\cN=1$ curves is \cite{Kapustin:1996nb,Kitao:1996mb,Giveon:1997gr,Csaki:1997zg,Gremm:1997sz,Lykken:1997gy,Giveon:1997sn,deBoer:1997zy,Burgess:1998jh,Csaki:1998dp,Hailu:2002bg,Hailu:2002bh}. Note that there are much more papers which studied the confining vacua using Riemann surfaces, which are not the $\cN=1$ curves in the sense used here. }

We begin by studying $\cN=1$ $\SU(2)^3$ gauge theory coupled to one trifundamental chiral multiplet, $Q_{\alpha_1\alpha_2\alpha_3}$. As we will see, the low energy dynamics can be described by one modulus field $U$ and two unbroken $\U(1)$ fields. We find that the curve is  given by  \begin{equation}
v^2 + U = F_{0,3}(z),
\end{equation} where $F_{0,3}(z)$ is a meromorphic \emph{function} on the sphere (i.e.~a genus-zero surface) with three poles. 
We then generalize it to a class of theories constructed according to a trivalent graph from $\SU(2)$ gauge groups and trifundamental multiplets.
We will see that this class of theories always have just one modulus field, and that the curve is given by \begin{equation}
v^2 + U = F_{g,n}(z),
\end{equation} where $F_{g,n}(z)$ is a meromorphic \emph{function} on a genus-$g$ Riemann surface with $n$ poles.
Here $g$ and $n$ is determined by the trivalent graph. 

The curves we find is similar to the ones for $\cN=2$ theories studied by Gaiotto in \cite{Gaiotto:2009we}, where the branch points of the double cover is given by meromorphic \emph{quadratic differentials} instead of meromorphic \emph{functions}. The curves are not very directly related, however, because the vacuum expectation values (vevs) of the trifundamentals were set to zero in \cite{Gaiotto:2009we}, while we study the phase where it is the trifundamentals which get the vev. 

The rest of the paper is organized as follows. In Sec.~\ref{sec:onetrifundamental}, our basic example, $\SU(2)^3$ gauge theory coupled to one trifundamental, is studied in detail. The monodromy on the moduli space is studied from various points of view, and the $\cN=1$ curve is determined there.  Then in Sec.~\ref{sec:general} we present the generalization to a class of $\cN=1$ theories associated to trivalent graphs, constructed from $\SU(2)$ gauge groups and trifundamentals. We propose the $\cN=1$ curves for these theories, and provide many checks. In some models we find runaway behavior.  We conclude the paper with discussions in Sec.~\ref{sec:conclusions}. In an Appendix we discuss how the flavor symmetry charge of a monopole can be induced via the Wess-Zumino-Witten terms, which is important to understand the monodromy on the moduli space.

\section{SU(2)$^{\mathbf{3}}$ with a trifundamental}\label{sec:onetrifundamental}

In this section, we consider ${\cal N}=1$  $\SU(2)_1 \times \SU(2)_2 \times \SU(2)_3$ gauge theory with a chiral multiplet 
$Q_{\alpha_1\alpha_2\alpha_3}$ in the  trifundamental representation with zero superpotential.
Here $\a_i=1,2$ is an index of the gauge group $\SU(2)_i$. 
We denote the dynamical scale of the gauge group $\SU(2)_i$ by $\Lambda_i$.
Our aim is to study the low energy dynamics of the model, 
which is described by $\U(1) \times \U(1)$ gauge fields and one moduli field as we will see.

\subsection{The moduli space and the semi-classical monodromy}\label{sec:qualitative}
We start from a classical analysis.
The gauge invariant moduli field can be constructed as follows. We  first define triplets of $\SU(2)_i$
\begin{equation}
A_{\a_1}^{~\b_1} =\frac{1}{2}Q_{\alpha_1\alpha_2\alpha_3}Q^{\beta_1\alpha_2\alpha_3}, \quad
B_{\a_2}^{~\b_2} =\frac{1}{2}Q_{\alpha_1\alpha_2\alpha_3}Q^{\alpha_1\beta_2\alpha_3}, \quad
C_{\a_3}^{~\b_3} =\frac{1}{2}Q_{\alpha_1\alpha_2\alpha_3}Q^{\alpha_1\alpha_2\beta_3}. \label{eq:triplet}
\end{equation}
where gauge indices are raised and lowered by $\varepsilon_{\a\b}$.
We then define 
\beq
U^{(1)}=\frac{1}{2} A_{\a_1}^{~\b_1}A_{\b_1}^{~\a_1},~~~U^{(2)}=\frac{1}{2} B_{\a_2}^{~\b_2}B_{\b_2}^{~\a_2},
~~~U^{(3)}=\frac{1}{2} C_{\a_3}^{~\b_3}C_{\b_3}^{~\a_3}. \label{eq:threedefofmoduli}
\eeq
At the classical level, these fields are the same,
\beq
U \equiv U^{(1)}=U^{(2)}=U^{(3)}.
\eeq
and hence we can parametrize the moduli space by $U$. $U$ is known as  Cayley's hyperdeterminant of $Q_{\alpha_1\alpha_2\alpha_3}$. 

Up to gauge transformation, the solution to the D-term equation is given by 
\beq
\vev{Q_{111}}=\vev{Q_{222}}=u,~~~{\rm other}~\vev{Q_{\a_1\a_2\a_3}}=0.
\eeq
Then $U=u^4/4$.  The gauge group is spontaneously broken down to 
$\U(1) \times \U(1)$, described by the gauge fields $A^{(i)}$ of $\U(1)_i \subset \SU(2)_i$  with a constraint
\beq
A^{(1)}+A^{(2)}+A^{(3)}=0.
\eeq

Next, let us study the semi-classical behavior. 
The one-loop gauge coupling of the gauge group $\SU(2)_i$ at the energy scale $u$, where the gauge groups are broken to $\U(1)$,
is given by
\beq
\tau_i =\frac{4\pi i}{g_i^2}+\frac{\theta_i}{2\pi} \simeq \frac{i}{2\pi}\log \left(\frac{U}{\Lambda^4_i}\right).
\eeq
Then, by taking the basis of the low energy $\U(1) \times \U(1)$ gauge fields as $A^{(2)}$ and $A^{(3)}$ with $A^{(1)}=-A^{(2)}-A^{(3)}$,
the effective gauge coupling of the low energy theory ${\bm \tau}=(\t_{ij})_{2 \leq i,j \leq 3}$
\beq
{\bm \t} \simeq
2\left(
\begin{array}{cc}
\t_2+\t_1 & \t_1 \\
\t_1 & \t_3+\t_1
\end{array}
\right).
\eeq
Here we normalized the couplings so that a doublet has charge $\pm1$, as in 
 Ref.~\cite{Seiberg:1994aj,Intriligator:1994sm}.
Therefore, the monodromy \begin{equation}
M=\begin{pmatrix}
\mathbf{A}&\mathbf{B}\\
\mathbf{C}&\mathbf{D}
\end{pmatrix}
\end{equation} acting on $\bm{\tau}$ as 
\beq
{\bm \t} \to ({\mathbf  A}{\bm \t}+{\mathbf B})({\mathbf C}{\bm \t}+{\mathbf D})^{-1}
\eeq
around the large circle in the $U$ plane  is given by 
\beq
M_{\infty}=-\left(
\begin{array}{cc}
{\mathbf 1}&{\mathbf B}_{\infty} \\
{\mathbf 0}&{\mathbf 1}
\end{array}
\right),
\qquad
{\mathbf B}_{\infty}=\left(
\begin{array}{cc}
-4&-2\\ 
-2&-4
\end{array}
\right). \label{eq:weakmonodromy}
\eeq 
Note the overall minus sign. Going around the $U$-plane once corresponds to sending $u$ to $iu$. 
This corresponds to performing the Weyl reflection $(
\begin{smallmatrix}
0&i\\
i&0
\end{smallmatrix}) \in \SU(2)$ 
for all three $\SU(2)$ gauge groups, which reverses the sign of the charges under the unbroken $\U(1)^2$ subgroup.

\subsection{Strong-coupling monodromy}
\begin{figure}
\begin{center}
\includegraphics[width=.55\textwidth]{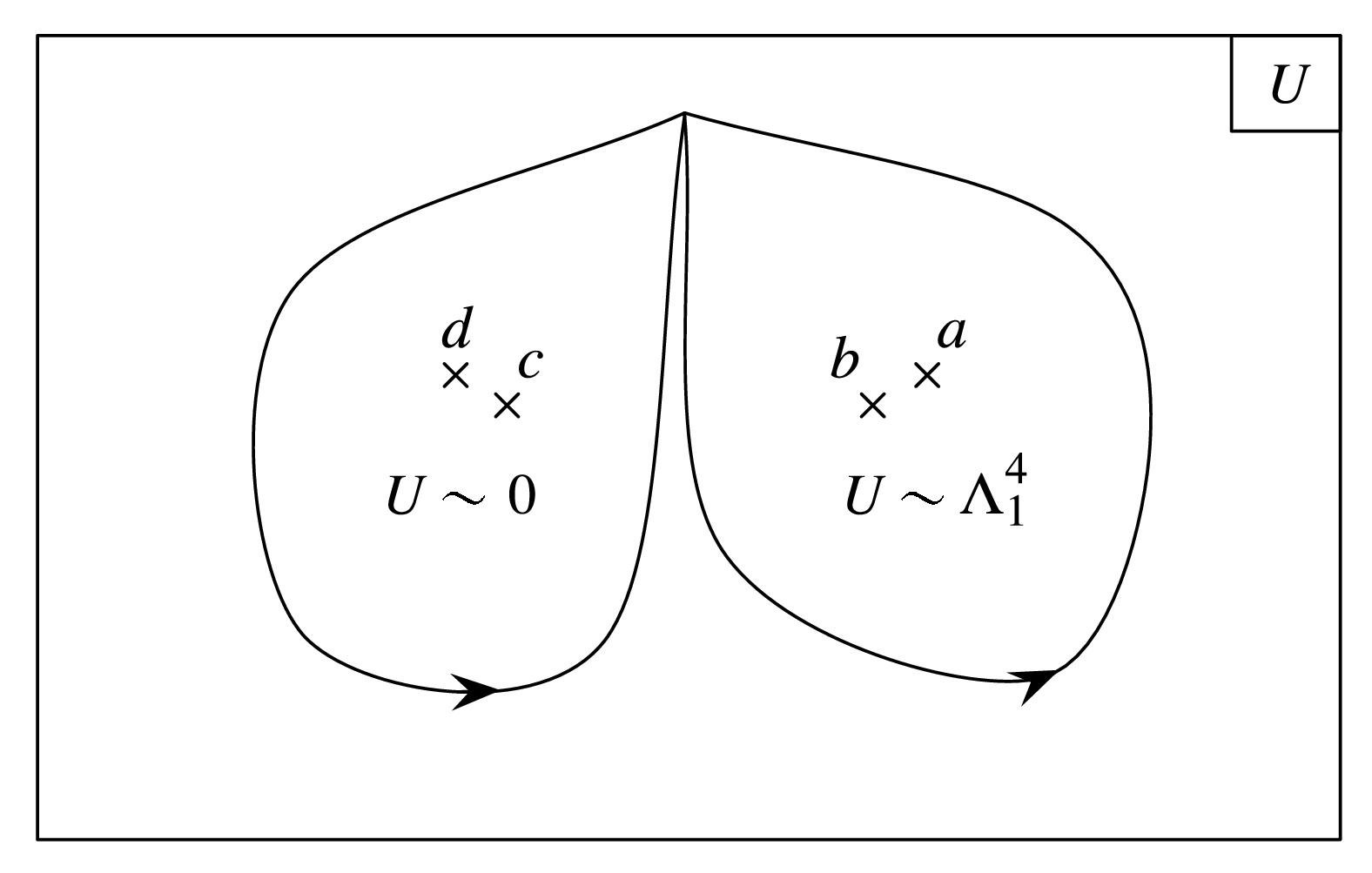}
\caption{The $U$-plane in the limit $\Lambda^4_2,\Lambda^4_3 \ll \Lambda^4_1$. 
There are four singular points in the $U$-plane where massless charged particles (monopole or dyon) appear.
We denote two singular points close to $U\sim\Lambda_1^4$ as $a$ and $b$,
and those close to $U\sim 0$ as $c$ and $d$.}
\label{fig:1}
\end{center}
\end{figure}

Now we consider the dynamics of the model at the strong-coupling region. 
When $\Lambda^4_2=\Lambda^4_3=0$~\cite{Seiberg:1994bz},
the quantum theory is described by the deformed moduli space,
which in terms of the variables defined above is given by
\beq
U^{(3)}-U^{(2)}=\Lambda^4_1. \label{eq:deformedmoduli}
\eeq
Let us turn on small but nonzero $\Lambda^4_1,~\Lambda^4_2 \ll \Lambda^4_1$.
Then, the dynamics is described by $\SU(2)_2 \times \SU(2)_3$ gauge theory with the adjoint fields $B_{\a_2}^{~\b_2}$ and 
$C_{\a_3}^{~\b_3}$ of each gauge group
constrained by Eq.~(\ref{eq:deformedmoduli}). The gauge group is broken down to $\U(1) \times \U(1)$ at generic points of the moduli space.
The $\SU(2)_i$ gauge group ($i=2,3$) is restored at the point $U^{(i)}=0$. 
Near this point, the dynamics is described by the $\SU(2)_i$ gauge group with an adjoint field, 
and quantum mechanically the point $U^{(i)}=0$ is split into two points where 
a monopole or a dyon becomes massless as in the case of the pure ${\cal N}=2$ Yang-Mills theory~\cite{Seiberg:1994rs}. 
Therefore, in the moduli space described by $U \equiv U^{(3)} $,
there are four singular points where massless charged particles appear, see Figure~\ref{fig:1}.
We denote singular points  close to $U\sim \Lambda_1^4$ by $a$, $b$, and
those close to $U\sim 0$ by  $c$, $d$.

Let us take the two unbroken $\U(1)$ gauge groups as $\U(1)_i \subset \SU(2)_i~(i=2,3)$ with the gauge fields $A^{(i)}$, and consider the
monodromies around the points $U \sim 0$ and $U \sim \Lambda^4_1$ as described in Figure~\ref{fig:1}. 
Naively, the monodromy around $U \sim 0$ is given by \begin{equation}
\tau_2\to\tau_2, \qquad \tau_3\to \tau_3 -4,
\end{equation} and similarly the monodromy around $U \sim \Lambda^4_1$ is
\begin{equation}
\tau_2\to\tau_2-4, \qquad \tau_3\to \tau_3.
\end{equation}
However, those monodromies alone do \emph{not} reproduce the weakly coupled monodromy $M_\infty$ in \eqref{eq:weakmonodromy}.

The missing piece of information is the fact \cite{Intriligator:1995id} that the monopole of $\SU(2)_2$ becomes a doublet of $\SU(2)_3$ due to the existence \cite{Manohar:1998iy} of the Wess-Zumino-Witten term on the moduli space. We discuss the process in detail in Appendix \ref{app:A}. 

Let us denote electric and magnetic charges under  $A^{(2,3)}$ of a particle by  ${}^t(e_2,e_3;m_2,m_3)$.
Then the charges of particles becoming massless at  $a$, $b$, $c$, $d$ have charges  \begin{equation}
q_a={}^t(0,\rho;1,0), \quad q_b={}^t(2,\rho;1,0); \quad
q_c={}^t(\sigma,0;0,1), \quad q_d={}^t(\sigma,2;0,1).
\end{equation} where $\rho$, $\sigma$ are the induced flavor charges, which will be determined below by demanding that $q_{a}$ is a doublet under $\SU(2)_3$, and that $q_{c}$ is a doublet under $\SU(2)_2$.

Monodromies produced by these particles can be calculated by going to the frame where each is purely electric, and are given by 
\begin{align}
M_a&=S_2 \begin{pmatrix}
\mathbf{1} & \mathbf{B}_{1,\rho}\\
\mathbf{0} & \mathbf{1}
\end{pmatrix} S_2^{-1}, &
M_b&= T_2^2 M_a T_2^{-2},\\
M_c&=S_3 \begin{pmatrix}
\mathbf{1} & \mathbf{B}_{\sigma,1}\\
\mathbf{0} & \mathbf{1}
\end{pmatrix} S_3^{-1}, &
M_d&=T_3^2 M_c T_3^{-2}.
\end{align} Here, $\mathbf{B}_{p,q}=\begin{pmatrix}
p^2 & pq \\
pq & q^2
\end{pmatrix}$, 
$S_2$ and $T_2$ maps $(e_2,m_2)$ to $(-m_2,e_2)$  and 
 $(e_2+m_2,m_2)$ without changing $(e_3,m_3)$, and similarly for $S_3$ and $T_3$.

The $\SU(2)_3$ charge of $q_a$ can be read off from the $\SU(2)_3$ Weyl reflection,
which can be performed by transporting $q_a$ around points $c$ and $d$. One finds  \begin{equation}
M_cM_d {} q_a = q_a', \quad
M_cM_d {} q_a' = q_a, 
\end{equation} 
where $q_a'-q_a={}^t(0,2(\sigma-\rho);0,0)$. 
$q_a$ should be a doublet, which fixes $\sigma-\rho=1$. Using the freedom of the basis change, we can set $\rho=0$, $\sigma=1$.  With $\rho$ and $\sigma$ determined, we find that  the monodromy around infinity is correctly reproduced:\begin{equation}
M_\infty=M_aM_bM_c M_d.
\end{equation}

\subsection{The curve}\label{sec:derivation}
In a supersymmetric theory,  the physical moduli-dependent couplings of the low energy $\U(1)$ fields 
is holomorphic, and can be conveniently encoded in the complex structure of a Riemann surface, which is called the $\cN=1$ curve. 
Note that it does not come with the Seiberg-Witten differential encoding the BPS masses when there is only $\cN=1$ supersymmetry.

The case $\Lambda^4_3=0$, i.e. the $\SU(2)_1 \times \SU(2)_2$ gauge theory
with two bifundamental fields was studied by Intriligator and Seiberg in ~\cite{Intriligator:1994sm}.
The low energy theory is a $\U(1)$ gauge theory with a moduli field $C_{\a_3}^{~\b_3}$ in the adjoint representation of the
flavor group $\SU(2)_3$. The curve  depends on the flavor-invariant combination of $C$, i.e. $U\equiv U^{(3)}=C^2$,
and was determined in \cite{Intriligator:1994sm} to be \begin{equation}
y^2=x(x^2+(\Lambda_1^4+\Lambda_2^4-U)x+\Lambda_1^4\Lambda_2^4).
\end{equation}
We use the following form 
\beq
y^2=Q_2(x)(P_2(x)-UQ_2(x))
\eeq where \beq
Q_2(x)=x, \quad
P_2(x)=\Lambda_1^4 x^2+(\Lambda^4_1-\Lambda^4_2)x-\Lambda^4_2.
\label{eq:intriseib}
\eeq 
instead, where we redefined  variables as $x\to \Lambda_1^4 x$, $y\to\Lambda_1^4 y$. We also changed the sign of $\Lambda_2^4$.\footnote{In the conventions in \cite{Intriligator:1994sm}, $U^{(3)}-U^{(1)}=\Lambda_2^4$ and 
$U^{(3)}-U^{(2)}=\Lambda_1^4$. In this paper, we prefer to treat $U^{(1,2,3)}$ and $\Lambda_{1,2,3}^4$ invariant under cyclic permutation. Therefore we redefine $\Lambda_2^4$ by a minus sign so that $U^{(1)}-U^{(3)}=\Lambda_2^4$ instead. }

Now let us consider the $\SU(2)^3$ theory. 
The low energy gauge group is $\U(1) \times \U(1)$, and hence we expect that the curve is a genus-two Riemann surface, 
which can always be represented by  a hyperelliptic curve of the form
\beq
y^2=f_6(x,U,\Lambda^4_i), \label{eq:abstractcurveform}
\eeq
where $f_6$ is a polynomial of $x$ of degree 6 (or degree 5 if one of the zero points of $f_6$ is taken at infinity.)

As discussed in the previous subsection, there are four singular points in the $U$-plane. 
This means that the discriminant of the polynomial $f_6$ should be a degree four polynomial of $U$.
This is  a very strong constraint on the family of curves, 
because the discriminant of a generic $f_6$ is a polynomial of degree much larger than four.

Let us now derive our curve. 
We introduce one additional massive flavor of quarks 
$P_{\alpha_1 i}~(i=1,2)$ in the fundamental representation of $\SU(2)_1$,
with the mass term
\begin{eqnarray}
W=\frac{m}{2}P_{\alpha_1 i}P^{\a_1 i}.\label{eq:mass}
\end{eqnarray}

From the point of view of $\SU(2)_1$ gauge group,
the theory has two massless and one massive quarks in the doublet representation.
Let us denote the dynamical scale of $\SU(2)_1$ before integrating out the additional massive quarks
by $\Lambda_\text{high}$, which satisfies $\Lambda_1^4=m\Lambda_\text{high}^3$.

We consider the regime $\Lambda_{\rm high} \gg m \gg \Lambda_{2,3}$,
The  theory below the scale $\Lambda_\text{high}$ is described by composite fields 
\begin{equation}
X = \frac{1}{2}  P_{\alpha_1 i}P^{\a_1 i},  \qquad
Z_{\alpha_2\alpha_3i} =  Q_{\alpha_1\alpha_2\alpha_3} P^{\a_1}_{i}, 
\end{equation}
and $B_{\a_2}^{~\b_2}$ and $C_{\a_3}^{~\b_3}$ defined in Eq.~(\ref{eq:triplet}).
In terms of these fields, the dynamical superpotential  of $\SU(2)$ theory with three flavors
together with the mass term \eqref{eq:mass} is given by 
\begin{eqnarray}
W_{\rm eff}=\frac{1}{\Lambda_{\rm high}^3}\left[X(U^{(2)}-U^{(3)}+\Lambda_1^4)-\frac{1}{2}B_{\b_2}^{~\a_2}Z_{\alpha_2\alpha_3i}Z^{\beta_2\alpha_3i}  
+\frac{1}{2}C_{\b_3}^{~\a_3}Z_{\alpha_2\alpha_3i}Z^{\alpha_2\beta_3i}  \right]. \label{eq:effsuperpot}
\end{eqnarray}
In this case, there is no Wess-Zumino-Witten term \cite{Manohar:1998iy}, and we can analyze the system using this superpotential alone.

Note that $Z_{\alpha_2\alpha_3i}$ ($i=1,2$) are bifundamental fields under the gauge group $\SU(2)_2 \times \SU(2)_3$. Therefore, aside from the term proportional to $X$,
the theory is holomorphically equivalent as the ${\cal N}=2$ $\SU(2) \times \SU(2)$ quiver gauge theory with 
one  bifundamental hypermultiplets. The curve of this model was found by Witten in \cite{Witten:1997sc}
and has the form
\begin{eqnarray}
\tilde{\Lambda}_2^2x^3-(v^2+u^{(2)})x^2-(v^2+u^{(3)})x-\tilde{\Lambda}_3^2=0,
\end{eqnarray}
where $u^{(2,3)}$ is the vev of the triplets and $\tilde{\Lambda}_{2,3}^2$  are two dynamical scales of the model.
We can roughly identify 
\beq
(u^{(2)}, u^{(3)}, \tilde{\Lambda}_2^2 , \tilde{\Lambda}_3^2 ) \sim 
\frac{1}{\Lambda_\text{high}^2} (U^{(2)},U^{(3)}, \Lambda_2^4,\Lambda_3^4).
\eeq 
Imposing the constraint from the terms proportional to $X$,
the curve of our model in the regime $\Lambda_1^4 \gg \Lambda_{2,3}^4$ is given by 
\begin{eqnarray}
c\Lambda_2^4 x^3-(v^2+U-\Lambda_1^4)x^2-(v^2+U)x-c'\Lambda_3^4 \sim 0 \label{eq:roughcurve1}
\end{eqnarray}
where $U \sim U^{(3)}$.

Then, the curve for general $\Lambda_{1,2,3}^4$ should be given by 
\begin{eqnarray}
c\Lambda_2^4 x^3-(v^2+U-\Lambda_1^4-c''\Lambda_2^4)x^2-(v^2+U+c''' \Lambda_3^4)x-c'\Lambda_3^8=0, \label{eq:undeterminedcurve}
\end{eqnarray} where we used the possibility to shift $U$ by a dimension-4 quantity.
This should reduce to the curve \eqref{eq:intriseib} when $\Lambda_2\to0$ or $\Lambda_3\to 0$.
This fixes $c=c'=c''=c'''=1$. 
Redefining $y=x(x+1)v$,
we find that the curve is given by 
\beq
y^2=x(x+1)\left[
(\Lambda_2^4 x^3+ (\Lambda_1^4+\Lambda_2^4)x^2-\Lambda_3^4x-\Lambda_3^4)
-Ux(x+1)
\right].
\label{eq:exactcurve}
\eeq

\subsection{Monodromy from the curve}

The curve we obtained has the general form
\beq
y^2 = f_6(x)=Q_3(x)\left(P_3(x)-UQ_3(x)\right),  \label{eq:degreesixpolynomial}
\eeq
where $P_3$ and $Q_3$ are polynomials of $x$ of degree three which are independent of $U$,
\begin{align}
Q_3(x) &= (x-a)(x-b)(x-c),  &
P_3(x) &= (x-d)(x-e)(x-f).
\end{align}
The monodromy can be studied more easily from this general perspective; 
$(a,b,c)$ can be later mapped to $(0,1,\infty)$ by a conformal transformation, resulting in \eqref{eq:exactcurve}.

\begin{figure}
\begin{center}
\includegraphics[scale=0.15]{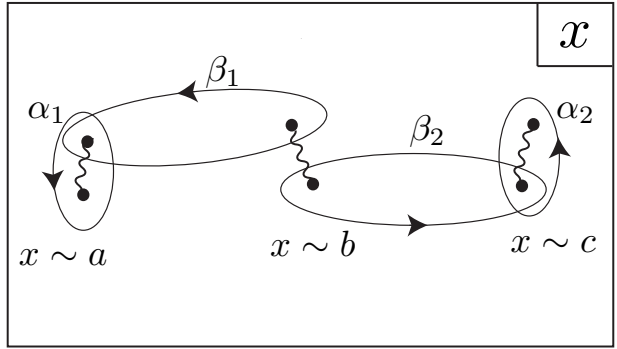}
\caption{Cycles $\a_1,\a_2,\b_1$ and $\b_2$ in the curve (\ref{eq:degreesixpolynomial}). Wavy lines represent the branch cuts between the zero points of $f_6$.}
\label{fig:2}
\end{center}
\end{figure}

When $U$ is finite, the three zeros of $P_3(x)-UQ_3(x)$ can never coincide with those of $Q_3(x)$. 
Therefore, the discriminant of $f_6$ is  the discriminant of $P_3(x)-UQ_3(x)$, which is a degree-four polynomial of $U$.
At the four finite values of $U$ at which $P_3(x)-UQ_3(x)$ has double zeros, 
one of the cycles of the curve pinches, producing the monodromy conjugate to $M_{a,b,c,d}$.  
When $U$ is very large, the zeros of $f_6$ are approximately given by
\beq
x ~\simeq~ a,~~b,~~c;~~a+{k_a}/{U},~~b+{k_b}/{U},~~c+{k_c}/{U} \label{eq:zeropointsoff6}
\eeq 
where $k_{a,b,c}$ are constants. 
We introduce branch cuts between $a$ and $a+k_a/U$, etc., and introduce cycles $\a_1,\a_2,\b_1$ and $\b_2$
in the curve (\ref{eq:degreesixpolynomial}) as shown in Figure~\ref{fig:2}. 
The monodromy of these cycles when $U$ is moved around $U\sim\infty$ 
can then be easily found, and we find 
\beq
\left(
\begin{array}{c}
\b_1 \\ \b_2 \\ \a_1 \\ \a_2
\end{array}
\right)
\to 
-\left(
\begin{array}{cccc}
1& 0 & -4 & -2 \\
0& 1 & -2 & -4 \\
0& 0 & 1 & 0\\
0& 0 & 0 & 1 
\end{array}
\right)
\left(
\begin{array}{c}
\b_1 \\ \b_2 \\ \a_1 \\ \a_2
\end{array}
\right).
\eeq
This reproduces $M_{\infty}$. Note that the overall minus sign comes from the exchange of branches $\pm y$ when $U$ is rotated once.

\subsection{Dynamics in the special loci}
Before starting the generalization, let us study special loci in the moduli space 
where interesting dynamics can be found. 
First, we can tune the parameters to realize $\cN=2$ $\SU(2)$ theory with one doublet hypermultiplet,
up to a holomorphic redefinition of fields. 
Recall that monopoles of the $\SU(2)_2$ are doublets of the $\SU(2)_3$ group.
The curve for the case $\Lambda_3^4=0$, \eqref{eq:intriseib}, implies that they
become massless when $\Lambda_1^4+\Lambda_2^4=0$.
Suggested by these considerations, 
let us write \begin{equation}
x=-\epsilon / x', \quad
y=\epsilon^2 y' \Lambda_2^2/x'{}^2,\quad
U=\epsilon^2 U' \Lambda_2^4,\quad  
\Lambda_1^4 +\Lambda_2^4 = \epsilon \mu \Lambda_2^4
\end{equation} where $\epsilon^3=\Lambda_3^4/\Lambda_2^4$,
and take the limit $\epsilon\to 0$ keeping the primed variables finite.
 Eq.~(\ref{eq:exactcurve}) becomes
\begin{equation}
y'^2=x'^3-U'x'^2-\mu x'+1.
\end{equation}
This is precisely the form of the curve of ${\cal N}=2$, $\SU(2)$ one flavor model studied in \cite{Seiberg:1994aj}.

In general, there are four singularities on the $U$-plane. Two out of the four collide when \begin{equation}
(\Lambda_1^4+\Lambda_2^4+\Lambda_3^4)^3- 27\Lambda_1^4\Lambda_2^4\Lambda_3^4=0.\label{eq:ADlocus}
\end{equation}
When this happens, mutually non-local particles become simultaneously massless, and we expect to have a nontrivial superconformal theory. 
When $\Lambda_{1,2,3}^4$ are in the region studied in the previous paragraph so that the theory is almost the $\cN=2$ theory with a massive doublet, the condition \eqref{eq:ADlocus} gives the locus where the Argyres-Douglas  superconformal field theory of $\cN=2$ $\SU(2)$ theory with one flavor \cite{Argyres:1995jj,Argyres:1995xn} is realized.
Our theory is then an $\cN=1$ exactly marginal perturbation of this theory.
However, as it is unlikely to have such an exactly marginal perturbation, 
we expect to have a symmetry enhancement to $\cN=2$ on the locus \eqref{eq:ADlocus}.

\section{Generalizations}\label{sec:general}

Our curve \eqref{eq:exactcurve}, \eqref{eq:degreesixpolynomial} can be recast in the form
\begin{equation}
v^2+U=F(z), \quad \text{where}\quad
F(z)=\frac{P_3(z)}{Q_3(z)}=-\Lambda_2^4z+\frac{\Lambda_1^4 z}{z-1}+ \frac{\Lambda_3^4}{z}.
\label{eq:3puncturecurve}
\end{equation} by the redefinition $v=y/Q_3(x)$ and $z=-x$.
We can regard $F$ as a meromorphic \emph{function} on a Riemann sphere with coordinate $z$
with poles  at $z=0,1,\infty$.  The residues are proportional to the dynamical scales. 
The curve \eqref{eq:intriseib} of $\SU(2)^2$ theory coupled to a bifundamental is obtained by taking the limit $\Lambda_3\to 0$.  Now $F$ has two poles on the sphere. 

This structure can be compared to that of the class of curves studied by Gaiotto for $\cN=2$ theories \cite{Gaiotto:2009we}, where the curve is given by $v^2=\phi_2(z)$  where $\phi_2(z)$ 
was a meromorphic \emph{quadratic differential} instead.
In this section we pursue this similarity and propose the curves 
for a large class of ${\cal N}=1$ gauge theories,
and perform many consistency checks of the proposal.

\begin{figure}
\begin{center}
\includegraphics[scale=0.35]{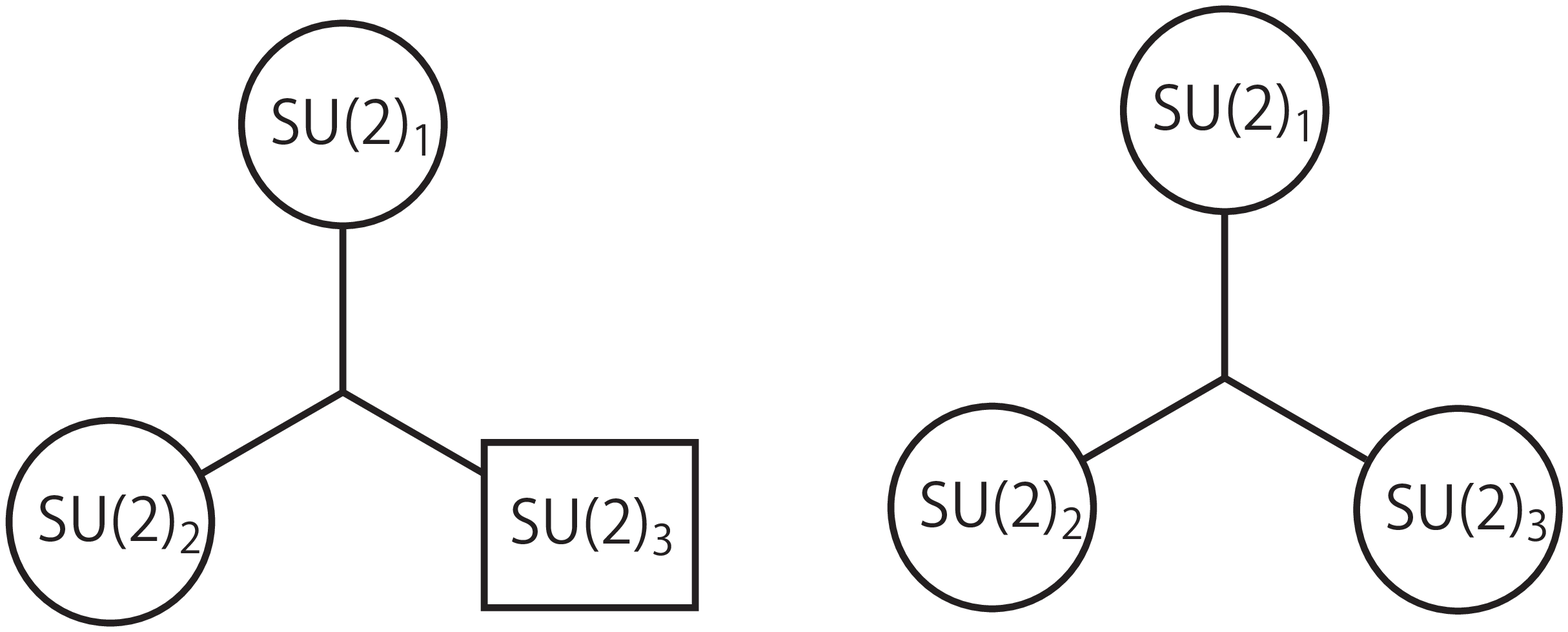}
\caption{Left: the model studied in \cite{Intriligator:1994sm}. 
Right: the model studied in section~\ref{sec:onetrifundamental}. Circles represent $\SU(2)$ gauge groups, boxes represent
global $\SU(2)$ symmetry groups, and trivalent vertices 
represent trifundamental fields. In this model, $g=0$ and $n=2$ (left) or $n=3$ (right).\label{fig:3}
}
\vskip 10mm
\end{center}

\begin{center}
\includegraphics[scale=0.25]{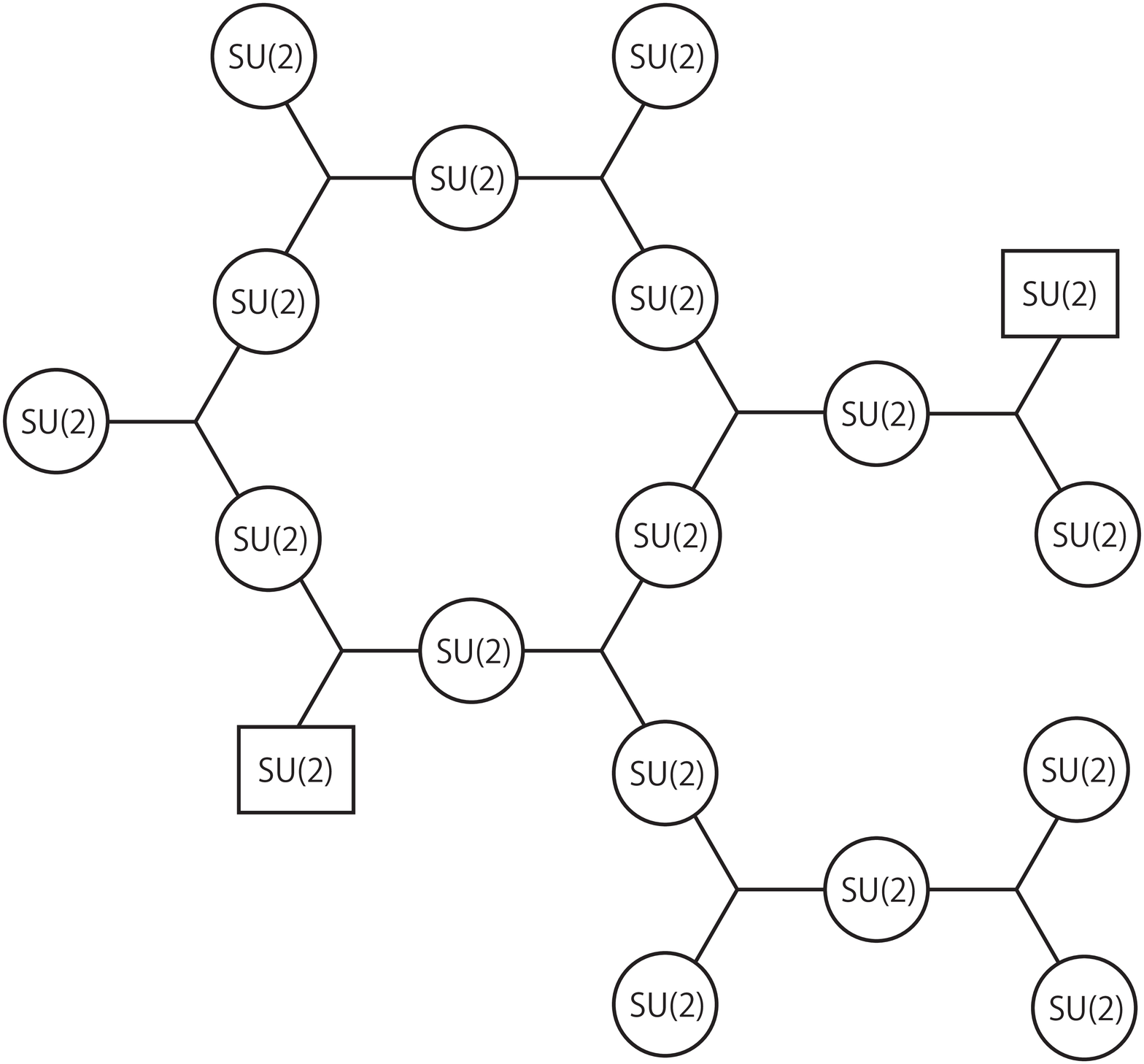}
\caption{An example of a generalized model. Circles represent $\SU(2)$ gauge groups, boxes represent
global $\SU(2)$ symmetry groups, and trivalent vertices 
represent trifundamental fields. In this example, $g=1$ and $n=7$.}
\label{fig:4}
\end{center}
\end{figure}

\subsection{The model}\label{sec:conjecture}

We consider the following class of $\cN=1$ theories, specified by a graph consisting of trivalent vertices connecting
circles or boxes, such that a circle can connect at most two vertices, and a box at most one vertex. 
The Lagrangian of the $\cN=1$ theory associated to a graph is given as follows:
\begin{itemize}
\item Label vertices by $v$, edges by $i$. Each edge has either a circle or a box. 
\item To each box $b$,  we have a flavor symmetry  $\SU(2)_b$.
\item To each circle $c$, we introduce a dynamical $\cN=1$ gauge multiplet $\SU(2)_c$.
\item To each trivalent vertex $v$, we introduce a trifundamental chiral multiplet $Q^{(v)}_{\alpha_i\alpha_j\alpha_k}$, 
where the vertex $v$ connects the edges $i,j$ and $k$.
\item An $\SU(2)_e$ multiplet coupled to only one trifundamental is called a \emph{leaf}. It effectively has two flavors, and  has the instanton factor $\Lambda^4_e$ associated to it.
\item An $\SU(2)_i$ multiplet coupled to two trifundamentals $Q^{(v)}$, $Q^{(v')}$ is called a \emph{stem}. 
For each  stem $\SU(2)_i$, we add a massive adjoint chiral multiplet $\Phi_i$ so that the $\SU(2)_i$ is semi-classically conformal. We call the UV coupling $\tau_i$. We then introduce the superpotential
\begin{equation}
W_i =m_{i} \Phi^{(i)}{}^{\alpha_i\beta_i}\Phi^{(i)}_{\alpha_i\beta_i}+ 
 \Phi^{(i)}{}^{\alpha_i\beta_i}(M^{(v,i)}_{\alpha_i\beta_i} +M^{(v',i)}_{\alpha_i\beta_i}) \label{eq:generalsuperpot}
\end{equation}  where \begin{equation}
M^{(v,i)}_{\alpha_i\beta_i}=Q^{(v)}_{\alpha_i\alpha_j\alpha_k}Q^{(v)}{}_{\beta_i}{}^{\alpha_j\alpha_k}.\label{eq:meson}
\end{equation} 
Note that we \emph{do not} add massive adjoints to the leaf $\SU(2)$s.
\end{itemize}

Then, $\SU(2)^2$ theory coupled to two bifundamentals studied by Intriligator and Seiberg in \cite{Intriligator:1994sm} 
and   $\SU(2)^3$ theory coupled to one trifundamental we studied in the previous section 
is defined by the graphs shown in Figure~\ref{fig:3}. 
A more complicated  graph is shown in Figure~\ref{fig:4}.

When there is no leaf dynamical $\SU(2)$ gauge groups and all $m_i=0$, 
our class of theories is exactly the class of $\cN=2$ theories studied by Gaiotto in \cite{Gaiotto:2009we}, where the curve on the Coulomb branch was also determined. 
Note however that in \cite{Gaiotto:2009we}, 
the vevs were given to $\Phi_i$, whereas we give vevs $Q^{(v)}$ and set $\vev{\Phi_i}=0$.
In a sense, we study the dynamics on the Higgs branch in the terminology of \cite{Gaiotto:2009we}.

${\cal N}=2$ theories can be deformed by the mass terms of the adjoint scalars in the $\cN=2$ gauge multiplets, see e.g.~\cite{Leigh:1995ep}.
In particular, when the theory has no leaf dynamical $\SU(2)$ gauge groups, our class of models is exactly the ones   studied in \cite{Maruyoshi:2009uk}. In the former the superconformal theory at the origin was analyzed, and in the latter the structure of the Higgs branch was determined.

Our main difference is the addition of the leaf $\SU(2)$ gauge groups. This leads the theory to the Abelian Coulomb phase, and the coupling matrix should be encoded in the family of $\cN=1$ curve. 
Suppose that there are $g$ loops in the graph and  $n$ leaf $\SU(2)$ gauge fields, and that the system is in the Abelian Coulomb phase.
We will soon see that this class of theory only has one modulus $U$ invariant under the flavor symmetries.
We propose that  the $\cN=1$ curve $\Sigma$ for the model is given by
\begin{eqnarray}
v^2+U=F_{g,n}(z)  \label{eq:conjecturedcurve},
\end{eqnarray}
where $F_{g,n}(z)$ is a meromorphic \emph{function} on a genus-$g$ Riemann surface $C_g$ with simple poles at $n$ points.
Furthermore, the position of poles and the complex structure of $C_g$ encode the UV couplings $\tau_i$ of the stem $\SU(2)_i$ groups, and the residues of $F_{g,n}(z)$ at the poles  encode the dynamical scales $\Lambda_e^4$ of the leaf $\SU(2)_e$ gauge groups.
The aim of the rest of the paper is to study the field theory dynamics of the models and see the consistency of the curves (\ref{eq:conjecturedcurve}) with the field theory expectations.

\subsection{Field theory analysis}
\subsubsection{Classical}
As always, our first task is to study the classical dynamics. 
Consider a graph with no dynamical leaf $\SU(2)$ gauge group,
which was the model studied in \cite{Maruyoshi:2009uk}.
The gauge-invariant operators of this model can be identified with the operators parameterizing the Higgs branch of the $\cN=2$ model  before turning on non-zero $m_i$. 
This Higgs branch was studied in detail in \cite{Hanany:2010qu}.\footnote{This branch was called the Kibble branch, because there are in general $g$  $\U(1)$ $\cN=2$ vector multiplets left unbroken, whose coupling constants can be naturally identified with the complex structure of $C_g$. Correspondingly, there are $g$ neutral scalars in the $\cN=2$ vector multiplet which we can still  turn on, but these $g$ directions are lifted by the $\cN=1$ adjoint mass deformations. }
As demonstrated there, typical  gauge-invariant moduli fields of this model are  the mesons $M^{(e)}\equiv M^{(v,e)}$ as defined in \eqref{eq:meson}
transforming under the triplet of the $\SU(2)_e$  leaf flavor symmetry (where $v$ is the vertex connected to the leaf edge $e$), and the `baryon' \begin{equation}
B=\prod_v Q^{(v)}
\end{equation} where the product is over all of the vertices, so that $B$ transforms as a doublet under all leaf $\SU(2)$ flavor symmetries. 
The flavor symmetry singlet $U_e=(M_e)^2$  is independent of the leaf $e$.

When the number $n$ of the leaf $\SU(2)$ gauge groups is non-zero, the gauge invariant moduli fields are the subset of the ones in \cite{Hanany:2010qu} which are neutral under the gauged leaf $\SU(2)$ symmetry. 
These fields can be read off from the generating function (7.1) in \cite{Hanany:2010qu},
and they are generated by triplets $M_e$ for each leaf $\SU(2)_e$ flavor symmetry, and the singlet $U$ satisfying $U=(M_e)^2$ for all $e$.

Let us now determine the unbroken gauge group at the generic point of the moduli space. 
The F-term equations $M^{(v,i)}+M^{(v',i)}=0$ for all stem edges $i$ connecting two vertices $v$, $v'$ can be solved as follows. Assign a direction to each edge, and let $h(v,i)=1$, $t(v,i)=2$ if the vertex $v$ is at the head of the edge $i$,
and let $h(v,i)=2$, $t(v,i)=1$ if the vertex $v$ is at the tail of the edge $i$.
We then set, for each vertex $v$ connecting three edges $i$, $j$ and $k$,
\begin{eqnarray}
\vev{Q^{(v)}_{h(v,i)h(v,j)h(v,k)}}=(-1)^{h(v,i)+h(v,j)+h(v,k)}\vev{Q^{(v)}_{t(v,i)t(v,j)t(v,k)}}=u, \label{eq:classicalvevs}
\end{eqnarray}
and set other components to zero. 

Let us denote by $A^{(i)}$ the vector field of $\U(1)_i\subset\SU(2)_i$.
The massless modes are those which satisfy
\begin{eqnarray}
(-1)^{h(v,i)}A^{(i)}+(-1)^{h(v,j)}A^{(j)}+(-1)^{h(v,k)}A^{(k)} = 0, \label{eq:massivephotons}
\end{eqnarray} for all $v$ connecting $(i,j,k)$. 
This has the usual form of the current conservation in an electrical circuit at each vertex. 
Therefore, when there are $n\ge 1$ leaf dynamical $\SU(2)$ gauge groups
and $g$ loops in the graph, the number of massless $\U(1)$ is given by $N_{\U(1)}=n+g-1$.

\subsubsection{Quantum}\label{subsec:quantum}
Let us now study how many singularities there should be on the $U$-plane. 
We start from the genus zero case. When $n=1$, the leaf gauge group $\SU(2)_1$ couples to the baryon $B$ which is a doublet under $\SU(2)_1$. Thus $\SU(2)_1$ is broken completely.
Now, let us weakly gauge other leaf $\SU(2)_e$ symmetries, $e=2,\ldots,n$. 
For a general choice of $\Lambda_1^4\gg \Lambda_{2,\ldots,n}^4$,
we expect each $\SU(2)_e$ to be almost restored at a different point on the $U$-plane, $U\sim U_e$.
At each such point, we have  $\SU(2)_e$ couped to a triplet $M^{(i)}$,
thus splitting the singularity at $U\sim U_e$ into two.  
Therefore, we expect to have $2n-2$ singularities on the $U$-plane.

Models with higher genus can be obtained by choosing two $\SU(2)$ leaf flavor symmetries, say $\SU(2)_e$ and $\SU(2)_f$ of the lower-genus model, and coupling the diagonal combination of them to a dynamical $\SU(2)$ gauge field. Let us study this procedure carefully for the genus-one model.

\begin{figure}
\begin{center}

\includegraphics[width=.6\textwidth]{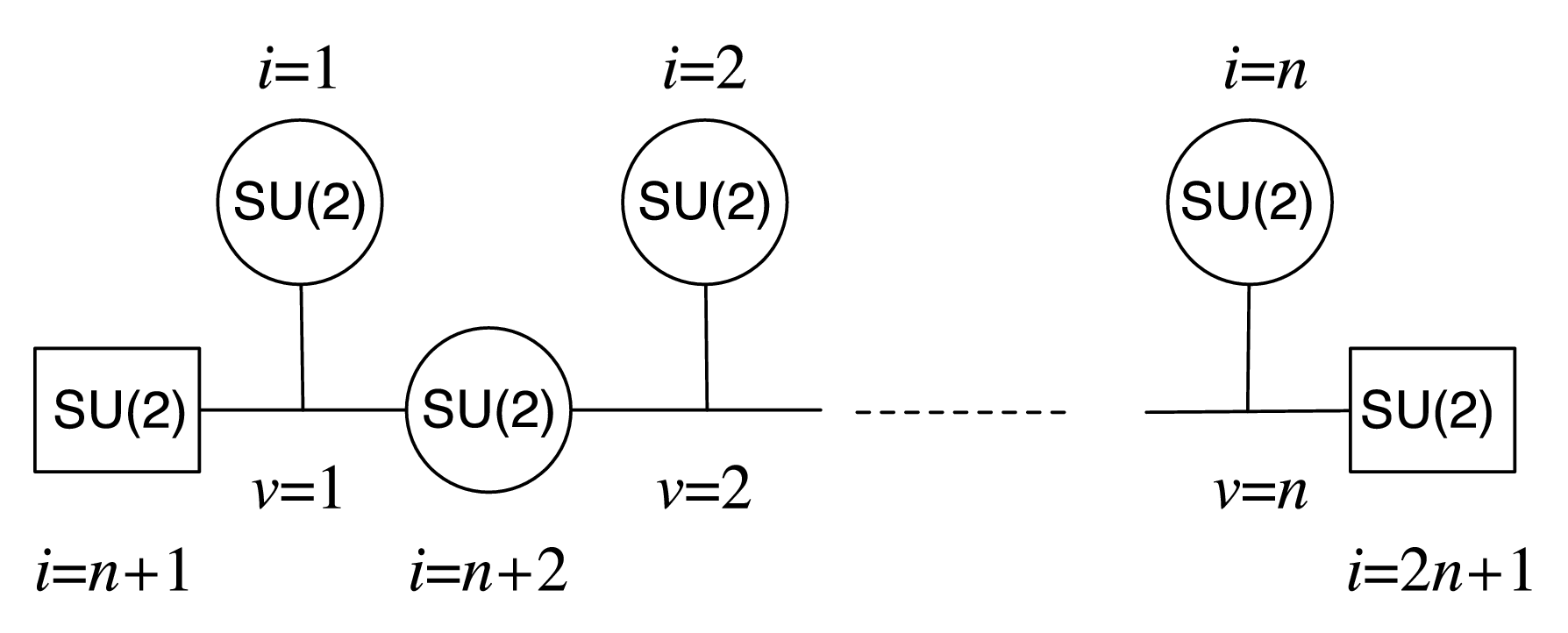}

\caption{A genus-one graph can be obtained by gauging the diagonal subgroup of $\SU(2)_{n}$ and $\SU(2)_{2n+1}$  of this linear graph.}
\label{fig:7}
\end{center}
\end{figure}

We consider the linear graph shown in Figure~\ref{fig:7}, where the vertices and the edges are named.
We consider the limit that all the scales $\Lambda_i^4$ ($i=1,\cdots,n$) are large enough so that we can go to a composite description.
The F-term conditions on the mesons as defined in \eqref{eq:meson} are 
\begin{equation}
M^{(i,n+i+1)} = -M^{(i+1,n+i+1)},  
\end{equation} for $i=1,\ldots,n-1$, and 
\begin{equation}
(M^{(i,n+i)})^2 =(M^{(i,n+i+1)})^2+ \Lambda_i'^4. 
\end{equation} for $i=1,\ldots,n$.
Here, we included possible quantum corrections from the stem $\SU(2)$ gauge interactions
as the definition of $\Lambda_i'^4$ in the right hand side.

Let us couple an $\SU(2)$ gauge group to the diagonal combination of $\SU(2)_{n+1}$  and $\SU(2)_{2n+1}$.
Effectively, this is an $\SU(2)$ gauge theory coupled to three triplets $\Phi$, $A\equiv M^{(1,n+1)}$ and $B\equiv M^{(n,2n+1)}$ with the superpotential \begin{equation}
W\sim X(A^2-B^2-\Lambda^4) +  \Phi (A +B) + m\Phi^2
\end{equation} where $X$ is a singlet Lagrange multiplier and 
$\Lambda^4=\sum_{i=1}^n \Lambda_i'^4. $
Writing $v=A+B$ and $u=A-B$, we have $W\sim X(uv-\Lambda^4)+\Phi v + m\Phi^2$.
$X$, $v$ and $\Phi$ can be integrated out, 
and we find the superpotential \begin{equation}
W\sim \frac{\Lambda^8}{mU}
\end{equation}
where $U=u^2$. Therefore, the vacuum runs away to infinity when $\Lambda^4\ne 0$.
When $\Lambda^4=0$, this $\SU(2)$ gauge group couples to one triplet $u$,
breaking $\SU(2)$  to $\U(1)$ and splitting the singularity at $U=0$  into two.

We find that the genus-1 model with $n$ leaf gauge symmetry is supersymmetric when 
\begin{eqnarray}
\sum_{i=1}^n \Lambda_i'^4=0 \label{eq:consistencycondition},
\end{eqnarray}
and that we expect $2n$ singularities on the $U$-plane.

A model with general genus $g$ is obtained by gluing $g$ pairs of $\SU(2)$ leaf flavor symmetries of the genus-zero model. 
Then, we expect that 
the supersymmetry requires one relation of the form  (\ref{eq:consistencycondition})  for each loop of the graph,
and that there are $2(n+g-1)=2N_{\U(1)}$ singularities on the $U$-plane.

\subsection{The curve}

Let us compare the qualitative behaviors we studied in the previous subsection
with the properties of the curve $\Sigma$ we propose.
Let us repeat the curve here: \begin{equation}
v^2+U=F_{g,n}(z) \label{eq:curves}
\end{equation} where $F_{g,n}(z)$ is a meromorphic function with $n$ poles on the Riemann surface $C$ of genus~$g$.
The complex structure of $C$ together with the positions of the poles specify the marginal couplings of the theory,
and the residues correspond to the dynamical scales of leaf $\SU(2)$ gauge groups.

\paragraph{Number of parameters}
Consider an R-symmetry with charge assignment
\begin{eqnarray}
Q^{(v)}:0,~~~\Phi^{(i)}:+2,~~~\Lambda_i^4:0,~~~\tau_i:0,~~~m_i:-2.
\end{eqnarray}
Since the curve is neutral under the R-symmetry,
the adjoint masses $m_i$ could only enter in the curve in the ratios $m_i/m_j$ to maintain the R-symmetry, i.e.
\begin{eqnarray}
F_{g,n}=F_{g,n}(z,\Lambda_i^4,\tau_i, m_i/m_j).
\end{eqnarray}
Now, let us take a limit $m_i \to 0$ with $m_i/m_j$ fixed. 
When $u \sim U^{1/4} \gg m_i$, all the fields charged under the low energy
$\U(1)$ fields have masses of order $u$, including the off-diagonal components of $\Phi_i$.
Thus, in this limit, the masses $m_i$ are irrelevant for the low energy dynamics of massless $\U(1)$
gauge interactions and hence $F_{g,n}$ does not depend on these mass parameters at all,
\begin{eqnarray}
F_{g,n} \to F_{g,n}(z,\Lambda_i^4,\tau_i). \label{eq:masslesslimitofF}
\end{eqnarray}
However, since we have fixed the ratio $m_i/m_j$ in the above limit and $F_{g,n}$ could depend only on these ratios, Eq.~(\ref{eq:masslesslimitofF}) means 
that $F_{g,n}$ cannot not depend on these ratios. Therefore, we conclude that $F_{g,n}$ is totally independent of the mass parameters $m_i$. 

There are $3g+n-3$  parameters in the complex structure moduli of the base curve together with the position of the poles. This equals the number of the marginal couplings $\tau_i$ of $\SU(2)_i$ stem gauge groups when all of the leaf $\SU(2)$ symmetries are gauged.

\paragraph{Uniqueness of the modulus}
Once the residues are specified, a meromorphic function can only be shifted by a constant on a Riemann surface. 
This agrees with the field theoretical fact that there is only one gauge- and flavor-invariant modulus, $U$.

\paragraph{Genus of the curve}
Let us compare the genus of the curve $\Sigma$ and $N_{\U(1)}$.
The curve $\Sigma(U)$ is a double cover of the genus $g$ Riemann surface $C_g$,
branched at the zeros and the poles of the meromorphic function $F_{g,n}(z)-U$.
As there are as many poles and zeros, there are $2n$ branch points in total.
Therefore, by the Riemann-Hurwitz formula, the genus of $\Sigma$ is 
\beq
g'=n+2g-1 = N_{\U(1)}+g.
\eeq
Therefore our family of curves describe $g$ additional $\U(1)$ fields than necessary, 
whose decoupling can be seen as follows.\footnote{Note that the genus of the $\cN=2$ Seiberg-Witten curve is always larger than the number of low energy $\U(1)$ fields, excepting the $\SU(N)$ linear quiver gauge theory. For example, the curve of pure $\cN=2$  $\SO(2n)$ comes with a $\mathbb{Z}_2$ action, and the field theory monodromy corresponds to the monodromy of the negative parity part under this $\mathbb{Z}_2$ action \cite{Brandhuber:1995zp}. For general $G$, a more general procedure is necessary, see e.g.~\cite{Martinec:1995by,Hollowood:1997pp}.}
Let us  pull back to $\Sigma(U)$ $2g$ cycles $\alpha_i$, $\beta_i$ and $g$ holomorphic differentials $\lambda_i$ on $C_g$, $(i=1,\ldots,g)$, and call them $\alpha^{+}_i$, $\beta^{+}_i$, $\lambda^{+}_i$.
On $\Sigma(U)$, one can choose $2(g'-g)$ cycles  $\alpha^{-}_j$, $\beta^{-}_j$ and $g'-g$ holomorphic differentials $\lambda^{-}_j$, $(j=1,\ldots,g'-g).$  
Superscripts on $\alpha,\beta,\lambda$ specify the parity of the object  under the exchange of two sheets, respectively. 
As such, they do not mix under the monodromy. 
As those with positive parity comes from the base curve $C_g$ which is independent of $U$,
the monodromy matrix is only nontrivial on the negative parity objects, which describe $g'-g=N_{\U(1)}$ Abelian gauge fields.

\paragraph{Singularities in the $U$-plane}

Let us compare the number of the singularities in the $U$-plane.  The $\cN=1$ curve becomes singular when $U$ is such that \begin{equation}
v^2=F_{g,n}(z)-U
\end{equation} has a double zero, say at $z_0$ on $C$.
Such a $z_0$ is a simple zero of $dF$. As $dF$ is a one-form with $n$ double poles, there are in general $2n+2g-2=2N_{\U(1)}$ simple zeros of $dF$. This agrees with the number determined in Sec.~\ref{subsec:quantum}.

\paragraph{Pinching of the curve}
Consider a graph with $g$ loops and $n$ leaf $\SU(2)$ gauge groups.
Take the limit where the marginal coupling of an stem $\SU(2)$ gauge group becomes zero.
In some cases this splits the graph into two, one with $g'$ loops and $n'$ leaf gauge groups,
the other with $g''$ loops and $n''$ leaf gauge groups, where
$g=g+g'$ and $n=n'+n''$. In the other cases the above limit reduces the number of loops from $g$ to $g-1$ with $n$ fixed.

In terms of the curve, the former process corresponds to the pinching of
the genus-$g$ surface $C$ to two surfaces $C'$ and $C''$, with genus $g'$ and $g''$, respectively.
The meromorphic function $F_{g,n}(z)$ 
with $n$ poles on $C$ then gives  $F_{g',n'}(z)$ on $C'$ and $F_{g'',n''}(z)$ on $C''$,
with $n=n'+n''$. In other words, the total number of the poles does not change,
agreeing with the trivial property on the field theory side. In the latter case the genus-$g$ surface pinches to genus-($g-1$) one with $n$ unchanged,
also agreeing with the field theory.
Here, it was important that $F(z)$ is a meromorphic \emph{function}. 
Otherwise, pinching generally introduces poles at the pinched point. 
\paragraph{Residues as the dynamical scales}
We  identify the residue at the poles with $\Lambda^4_i$ of the leaf $\SU(2)_i$ gauge group.
There are conditions on the residues when $g\ge 1$.
Let us first consider the case $g=1$.
We take a coordinate $z$ on the torus such that it satisfies the periodicity condition,
\begin{eqnarray}
z \cong z+l+m\tau,~~~l,m \in  {\mathbb Z}
\end{eqnarray}
where $\tau$ is the moduli parameter of the torus.
Then, $F_{1,n}(z)$ is given by
\begin{eqnarray}
F_{1,n}(z)=\sum_{l,m \in  {\mathbb Z}}\sum_{i=1}^n \frac{\tilde{\Lambda}_i^4}{z-c_i+l+m\tau} . \label{eq:genusonecurve}
\end{eqnarray}
For this series to converge,
we have to impose the condition
\begin{eqnarray}
\sum_{i=1}^n \tilde{\Lambda}_i^4=0. \label{eq:consistencycondition1}
\end{eqnarray}
This agrees with the constraint we saw from the field theory analysis, \eqref{eq:consistencycondition}.

For general $g$,
there are $g$ holomorphic one-forms $\lambda_i~(i=1,\cdots,g)$.
Then, we can construct $g$ meromorphic one-forms $F_{g,n} \cdot \lambda_i$. In general, 
the sum of residues of any meromorphic one-form in a compact Riemann surface
must vanish. Therefore we find $g$ linear constraints  on the residues of $F_{g,n}$,
as also seen from field theory analysis.~\footnote{
The Riemann-Roch theorem guarantees that there always is a rational function given the poles and the residues satisfying $g$ constraints when $n> 2g-2$, but this is not always the case when $n\le 2g-2$. It would be interesting to study how this point is reflected in the field theory side.}

\paragraph{Equivalence of the curves}

The number of poles of $F_{g,n}(z)$  depends on 
the number of leaf $\SU(2)$ gauge groups,
and not on the number of leaf $\SU(2)$ flavor symmetries.

\begin{figure}
\begin{center}
\includegraphics[height=.2\textwidth]{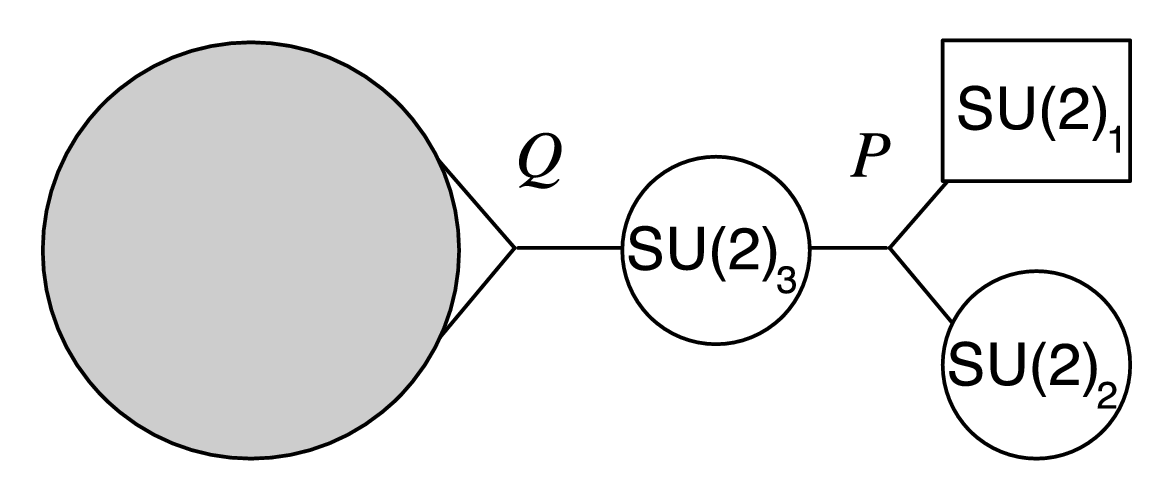} \qquad\quad
\includegraphics[height=.2\textwidth]{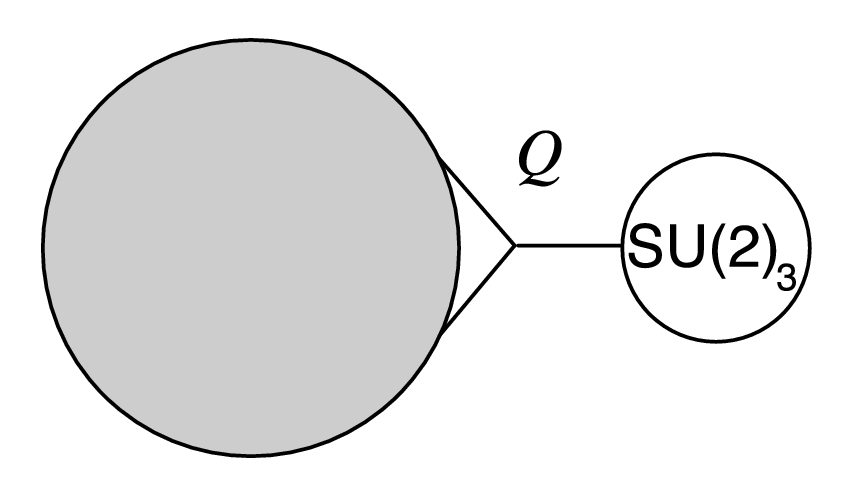}
\caption{A graph with an leaf flavor $\SU(2)_1$, and its decoupled form. The large shaded disk represents an arbitrary tree graph. $Q$ and $P$ are trifundamental fields.}
\label{fig:5}
\end{center}
\end{figure}

This independence can be understood by the following analysis,
at least when the leaf $\SU(2)$ flavor symmetry 
is at the place shown in Figure~\ref{fig:5}. 
The superpotential is given by
\begin{eqnarray}
W \sim W'+\Phi^{(3)}(PP+QQ)+m_{3} (\Phi^{(3)})^2 ,
\end{eqnarray}
where $W'$ is the superpotential of the remaining part denoted by the large shaded disk in the graph.
When  $\Lambda_2^4$ is large enough, we can go to a composite description.
Using $A=M^{(P,1)}$ and $B=M^{(P,3)}$,
the low energy superpotential is given by
\begin{eqnarray}
W \sim W'+\Phi^{(3)}(B+QQ)+m_{3} (\Phi^{(3)})^2+X\left(A^2-B^2-\Lambda_2^4 \right) ,
\end{eqnarray}
where $X$ is a Lagrange multiplier.
$\Phi^{(3)}$ and $B$ are massive and can be integrated out. Then, we obtain
\begin{eqnarray}
W \sim W'+X\left(A^2-U-\Lambda_2^4 \right),
\end{eqnarray}
where $U \sim Q^4$.
Then we have the almost decoupled theory as shown in the right of Figure~\ref{fig:5},
described by the superpotential $W'$,
and the $\SU(2)_3$ is now an leaf gauge group in this sector. 
The curve of this sector is the same as the original one. 
The flavor symmetry $\SU(2)_1$  acts only on the adjoint field $A$, constrained by the relation
\begin{eqnarray}
A^2=U+\Lambda_2^4. \label{eq:moduliconstraint}
\end{eqnarray}

\section{Conclusions}\label{sec:conclusions}
In this paper, we studied the Coulomb phase of a class of $\cN=1$ gauge theories composed of $\SU(2)$ gauge groups and trifundamental matter fields. We saw that its $\cN=1$ curve encoding the dependence of the low-energy coupling matrix of $\U(1)$ gauge fields is given by a double cover of a  Riemann surface, branched at zeros and poles of a meromorphic function on it.
We confirmed that the curve reproduces semi-classical behavior of the theory, and studied a few features of the strong-coupled dynamics using the curve thus obtained. 

We employed only field theoretical methods in this paper. 
But the close similarity of the curve we found and the curve in \cite{Gaiotto:2009we} suggests that the construction of the theory in terms of M5-branes should be possible, which explains the form of the curve. 
The brane construction of $\cN=1$ curves of theories with bifundamentals \cite{Giveon:1997sn,deBoer:1997zy} would be the obvious starting point.

Recalling that the trifundamental of $\SU(2)$ is the $T_2$ theory in the terminology of \cite{Gaiotto:2009we},
one generalization would be to replace $T_2$ theory by $T_N$ theory and  $\SU(2)$ gauge group by $\SU(N)$ gauge group in our construction.
We expect that we still find the theory in the Coulomb phase in the general case, 
but since the $T_N$ is a nontrivial superconformal theory there are many preliminary problems we need to solve.
For example, the contribution to the one-loop beta function from the $\SU(N)$ flavor symmetry of $T_N$ theory when coupled to a dynamical gauge field is the same as that of $N$ flavors of quarks. 
This is the value which causes the moduli space to deform. Then we first need to understand what it means to deform the moduli space of a nontrivial superconformal theory. 

In general, the Abelian Coulomb phase of $\cN=1$ gauge theories is not well studied, and deserves to be explored more extensively. For example, in this paper we found points in the moduli space where the Argyres-Douglas-type superconformal theory is realized.  In this case, we saw a strong indication that the infrared limit is indeed the $\cN=2$ Argyres-Douglas point of $\SU(2)$ gauge theory with one flavor, but we expect that there can be genuinely $\cN=1$ superconformal points with massless mutually non-local particles, embedded in the Coulomb phase.

\section*{Acknowledgements}
The authors are indebted to  Takuya Okuda and Yutaka Ookouchi for the collaboration at the early stage of the work; they are the ones who originally posed the problem studied in this paper. 
The authors thank Alexey Bondal for sharing his knowledge on the fibration of higher-genus Riemann surfaces on a sphere.
The authors also thank Kentaro Hori, Shoichi Kanno, Shigeki Sugimoto and Futoshi Yagi for discussions.
The authors are supported in part by World Premier International Research Center Initiative (WPI Initiative),  MEXT, Japan through the Institute for the Physics and Mathematics of the Universe, the University of Tokyo.
The work of YT also is supported in part by NSF grant PHY-0969448  and by the Marvin L. Goldberger membership through the Institute for Advanced Study.
The work of KY is supported in part by JSPS Research Fellowships for Young Scientists.

\appendix 

\section{Monopole flavor charge} \label{app:A}
Here we discuss how the monopole of an $\SU(2)$ theory can have an flavor symmetry under another global $\SU(2)$ symmetry
induced from a Wess-Zumino-Witten term. 

Recall the theory discussed in section~\ref{sec:qualitative}. 
We let $ \Lambda^4_1 \gg \Lambda^4_2$, and leave $\SU(2)_3$ ungauged. 
At the energy scale far below $\Lambda_1$, the dynamics is captured by the deformed moduli space (\ref{eq:deformedmoduli}).
The Wess-Zumino-Witten term plays an important role in this model~\cite{Manohar:1998iy}.

In the following the supersymmetry is not relevant, so let us consider a simpler model where the scalar fields live on $S^5$ \begin{equation}
(\Phi_I )^2 = (\phi_{a_2})^2 + (\varphi_{a_3})^2 = 1,
\end{equation} where $\Phi_I=(\phi_{a_2}, \varphi_{a_3})$ ($I=1,\ldots,6$, $a_2=1,2,3$, and $a_3=1,2,3$) are real fields, 
with $k$ units of  Wess-Zumino-Witten term \begin{equation}
S_{\mathrm{WZW}}  = \frac{2\pi  k}{\pi^3}
\int_X \frac{1}{5!}\epsilon_{IJKLMN} 
\Phi^I  d \Phi^J \wedge d\Phi^K \wedge d\Phi^L\wedge d\Phi^M\wedge d\Phi^N,
\end{equation} 
where, as always, $X$ is the five-dimensional ball whose boundary is the four-dimensional spacetime. 

This sigma model has $\SO(6)$ global symmetry. 
We couple dynamical  $\SO(3)_2 $ gauge fields to $\phi_{a_2}$,
and leave $\SO(3)_3 $ rotating $\varphi_{a_3}$ a global symmetry. 
Let us consider the vacuum where $\vev{\varphi_{a_3}}=0$. The dynamical $\SO(3)_2$ is broken down to $\U(1)$. Consider the 't Hooft-Polyakov monopole configuration.
The global $\SO(3)_3$ symmetry is still unbroken. We would like to determine how the monopole transforms under $\SO(3)_3$.

The important point is that both $\phi_{a_2}$ and $\varphi_{a_3}$ have a nontrivial profile: 
\begin{itemize}
\item At the asymptotic infinity, $\varphi_{a_3}\to 0$ and $\phi_{a_2}$ approaches the standard hedgehog configuration.
\item At the center of the monopole, the $\SO(3)_2$ symmetry should be restored and therefore $\phi_{a_2}\to 0$. Then $\varphi_{a_3}$ should have a vev there, 
$\varphi_{a_3}(r=0)\ne 0$. In other words, the $\SO(3)_3$ symmetry is broken in the interior of the monopole.
\end{itemize}

Therefore the monopole configuration comes with additional parameters 
$v_{a_3} \equiv \varphi_{a_3}(r=0)$ which satisfy $(v_{a_3})^2=1$, i.e.~the parameters live on $S^2$, acted by $\SO(3)_3$. 
The slow evolution of the monopole is then described by considering the motion $v_{a_3}(t)$ inside $S^2$, and 
we can show that the Wess-Zumino-Witten term reduces to $k$ units of effective magnetic flux inside this $S^2$. 

Let  the time direction be periodic. In that case $v_{a_3}(t)$ forms a closed orbit $\g$ in the $S^2$,
and we take a two-dimensional disk $D$ inside the $S^2$ with the boundary given by $\g$. Then we define a five dimensional manifold as
$X=D \times {\mathbb R}^3$, where ${\mathbb R}^3$ is the space dimensions. 
The integration over $\mathbb{R}^3$ can be easily done, and we obtain
\beq
\frac{k}{2} \int_{D} \frac{1}{2! } \epsilon_{a b c} v^{a}  d v^{b} \wedge d v^{c} .
\eeq
This is precisely the term representing the $k$ units of magnetic flux~\cite{Witten:1983tw}. 
Note in particular that if we change the integration region as $D\to S^2$, the above term gives $2 \pi k$.
The standard quantum mechanics on $S^2$ in the presence of magnetic flux tells us that the lowest energy modes transform under the spin $k/2$ representation of $\SO(3)_3$.
In our original supersymmetric situation,  we have $k=1$~\cite{Manohar:1998iy}. Therefore the monopole is a doublet under $\SO(3)_3$.

\bibliographystyle{utphys}
\bibliography{ref}

\providecommand{\href}[2]{#2}\begingroup\raggedright\begin{thebibliography}{10}

\bibitem{Intriligator:1994sm}
K.~A. Intriligator and N.~Seiberg, ``{Phases of ${\mathcal{N}}\!=1$
  Supersymmetric Gauge Theories in Four- Dimensions},''
  \href{http://dx.doi.org/10.1016/0550-3213(94)90215-1}{{\em Nucl. Phys.}
  {\bfseries B431} (1994) 551--568},
\href{http://arxiv.org/abs/hep-th/9408155}{{\ttfamily arXiv:hep-th/9408155}}.

\bibitem{Intriligator:1995id}
K.~A. Intriligator and N.~Seiberg, ``{Duality, Monopoles, Dyons, Confinement
  and Oblique Confinement in Supersymmetric SO(\hbox{$N_c$}) Gauge Theories},''
  \href{http://dx.doi.org/10.1016/0550-3213(95)00159-P}{{\em Nucl. Phys.}
  {\bfseries B444} (1995) 125--160},
\href{http://arxiv.org/abs/hep-th/9503179}{{\ttfamily arXiv:hep-th/9503179}}.

\bibitem{Argyres:1999fc}
P.~C. Argyres and A.~Buchel, ``{New S-Dualities in ${\mathcal{N}}\!=2$
  Supersymmetric $SU(2)$ $\times$ $SU(2)$ Gauge Theory},'' {\em JHEP}
  {\bfseries 11} (1999) 014,
\href{http://arxiv.org/abs/hep-th/9910125}{{\ttfamily arXiv:hep-th/9910125}}.

\bibitem{Gaiotto:2009we}
D.~Gaiotto, ``{${\mathcal{N}}\!=2$ Dualities},''
\href{http://arxiv.org/abs/0904.2715}{{\ttfamily arXiv:0904.2715 [hep-th]}}.

\bibitem{Maruyoshi:2009uk}
K.~Maruyoshi, M.~Taki, S.~Terashima, and F.~Yagi, ``{New Seiberg Dualities from
  ${\mathcal{N}}\!=2$ Dualities},''
  \href{http://dx.doi.org/10.1088/1126-6708/2009/09/086}{{\em JHEP} {\bfseries
  09} (2009) 086},
\href{http://arxiv.org/abs/0907.2625}{{\ttfamily arXiv:0907.2625 [hep-th]}}.

\bibitem{Benini:2009mz}
F.~Benini, Y.~Tachikawa, and B.~Wecht, ``{Sicilian Gauge Theories and
  ${\mathcal{N}}\!=1$ Dualities},''
  \href{http://dx.doi.org/10.1007/JHEP01(2010)088}{{\em JHEP} {\bfseries 01}
  (2010) 088},
\href{http://arxiv.org/abs/0909.1327}{{\ttfamily arXiv:0909.1327 [hep-th]}}.

\bibitem{Seiberg:1994rs}
N.~Seiberg and E.~Witten, ``{Monopole Condensation, and Confinement in
  ${\mathcal{N}}\!=2$ Supersymmetric Yang-Mills Theory},''
  \href{http://dx.doi.org/10.1016/0550-3213(94)90124-4}{{\em Nucl. Phys.}
  {\bfseries B426} (1994) 19--52},
\href{http://arxiv.org/abs/hep-th/9407087}{{\ttfamily arXiv:hep-th/9407087}}.

\bibitem{Seiberg:1994aj}
N.~Seiberg and E.~Witten, ``{Monopoles, Duality and Chiral Symmetry Breaking in
  ${\mathcal{N}}\!=2$ Supersymmetric QCD},''
  \href{http://dx.doi.org/10.1016/0550-3213(94)90214-3}{{\em Nucl. Phys.}
  {\bfseries B431} (1994) 484--550},
\href{http://arxiv.org/abs/hep-th/9408099}{{\ttfamily arXiv:hep-th/9408099}}.

\bibitem{Kapustin:1996nb}
A.~Kapustin, ``{The Coulomb Branch of ${\mathcal{N}}\!=1$ Supersymmetric Gauge
  Theory with Adjoint and Fundamental Matter},''
  \href{http://dx.doi.org/10.1016/S0370-2693(97)00209-8}{{\em Phys. Lett.}
  {\bfseries B398} (1997) 104--109},
\href{http://arxiv.org/abs/hep-th/9611049}{{\ttfamily arXiv:hep-th/9611049}}.

\bibitem{Kitao:1996mb}
T.~Kitao, S.~Terashima, and S.-K. Yang, ``{${\mathcal{N}}\!=2$ Curves and a
  Coulomb Phase in ${\mathcal{N}}\!=1$ SUSY Gauge Theories with Adjoint and
  Fundamental Matters},''
  \href{http://dx.doi.org/10.1016/S0370-2693(97)00261-X}{{\em Phys. Lett.}
  {\bfseries B399} (1997) 75--82},
\href{http://arxiv.org/abs/hep-th/9701009}{{\ttfamily arXiv:hep-th/9701009}}.

\bibitem{Giveon:1997gr}
A.~Giveon, O.~Pelc, and E.~Rabinovici, ``{The Coulomb Phase in
  ${\mathcal{N}}\!=1$ Gauge Theories with an LG-Type Superpotential},''
  \href{http://dx.doi.org/10.1016/S0550-3213(97)00297-6}{{\em Nucl. Phys.}
  {\bfseries B499} (1997) 100--124},
\href{http://arxiv.org/abs/hep-th/9701045}{{\ttfamily arXiv:hep-th/9701045}}.

\bibitem{Csaki:1997zg}
C.~Csaki, J.~Erlich, D.~Z. Freedman, and W.~Skiba, ``{${\mathcal{N}}\!=1$
  Supersymmetric Product Group Theories in the Coulomb Phase},''
  \href{http://dx.doi.org/10.1103/PhysRevD.56.5209}{{\em Phys. Rev.} {\bfseries
  D56} (1997) 5209--5217},
\href{http://arxiv.org/abs/hep-th/9704067}{{\ttfamily arXiv:hep-th/9704067}}.

\bibitem{Gremm:1997sz}
M.~Gremm, ``{The Coulomb Branch of ${\mathcal{N}}\!=1$ Supersymmetric
  $SU(\hbox{$N_c$}$) $\times$ $SU(\hbox{$N_c$}$) Gauge Theories},''
  \href{http://dx.doi.org/10.1103/PhysRevD.57.2537}{{\em Phys. Rev.} {\bfseries
  D57} (1998) 2537--2542},
\href{http://arxiv.org/abs/hep-th/9707071}{{\ttfamily arXiv:hep-th/9707071}}.

\bibitem{Lykken:1997gy}
J.~D. Lykken, E.~Poppitz, and S.~P. Trivedi, ``{Chiral Gauge Theories from
  D-Branes},'' \href{http://dx.doi.org/10.1016/S0370-2693(97)01220-3}{{\em
  Phys. Lett.} {\bfseries B416} (1998) 286--294},
\href{http://arxiv.org/abs/hep-th/9708134}{{\ttfamily arXiv:hep-th/9708134}}.

\bibitem{Giveon:1997sn}
A.~Giveon and O.~Pelc, ``{M Theory, Type IIA String and 4D ${\mathcal{N}}\!=1$
  SUSY $SU(N_L$) $\times$ $SU(N_R$) Gauge Theory},''
  \href{http://dx.doi.org/10.1016/S0550-3213(97)00687-1}{{\em Nucl. Phys.}
  {\bfseries B512} (1998) 103--147},
\href{http://arxiv.org/abs/hep-th/9708168}{{\ttfamily arXiv:hep-th/9708168}}.

\bibitem{deBoer:1997zy}
J.~de~Boer, K.~Hori, H.~Ooguri, and Y.~Oz, ``{K\"ahler Potential and Higher
  Derivative Terms from M Theory Five-Brane},''
  \href{http://dx.doi.org/10.1016/S0550-3213(98)00152-7}{{\em Nucl. Phys.}
  {\bfseries B518} (1998) 173--211},
\href{http://arxiv.org/abs/hep-th/9711143}{{\ttfamily arXiv:hep-th/9711143}}.

\bibitem{Burgess:1998jh}
C.~P. Burgess, A.~de~la Macorra, I.~Maksymyk, and F.~Quevedo, ``{Supersymmetric
  Models with Product Groups and Field Dependent Gauge Couplings},'' {\em JHEP}
  {\bfseries 09} (1998) 007,
\href{http://arxiv.org/abs/hep-th/9808087}{{\ttfamily arXiv:hep-th/9808087}}.

\bibitem{Csaki:1998dp}
C.~Csaki and W.~Skiba, ``{Classification of the ${\mathcal{N}}\!=1$
  Seiberg-Witten Theories},''
  \href{http://dx.doi.org/10.1103/PhysRevD.58.045008}{{\em Phys. Rev.}
  {\bfseries D58} (1998) 045008},
\href{http://arxiv.org/abs/hep-th/9801173}{{\ttfamily arXiv:hep-th/9801173}}.

\bibitem{Hailu:2002bg}
G.~Hailu, ``{${\mathcal{N}}\!=1$ Supersymmetric $SU(2)^r$ Moose Theories},''
  \href{http://dx.doi.org/10.1103/PhysRevD.67.085023}{{\em Phys. Rev.}
  {\bfseries D67} (2003) 085023},
\href{http://arxiv.org/abs/hep-th/0209266}{{\ttfamily arXiv:hep-th/0209266}}.

\bibitem{Hailu:2002bh}
G.~Hailu, ``{Quantum Moduli Spaces of Linear and Ring Mooses},''
  \href{http://dx.doi.org/10.1016/S0370-2693(02)03160-X}{{\em Phys. Lett.}
  {\bfseries B552} (2003) 265--272},
\href{http://arxiv.org/abs/hep-th/0209267}{{\ttfamily arXiv:hep-th/0209267}}.

\bibitem{Seiberg:1994bz}
N.~Seiberg, ``{Exact Results on the Space of Vacua of Four-Dimensional SUSY
  Gauge Theories},'' \href{http://dx.doi.org/10.1103/PhysRevD.49.6857}{{\em
  Phys. Rev.} {\bfseries D49} (1994) 6857--6863},
\href{http://arxiv.org/abs/hep-th/9402044}{{\ttfamily arXiv:hep-th/9402044}}.

\bibitem{Manohar:1998iy}
A.~V. Manohar, ``{Wess-Zumino Terms in Supersymmetric Gauge Theories},''
  \href{http://dx.doi.org/10.1103/PhysRevLett.81.1558}{{\em Phys. Rev. Lett.}
  {\bfseries 81} (1998) 1558--1561},
\href{http://arxiv.org/abs/hep-th/9805144}{{\ttfamily arXiv:hep-th/9805144}}.

\bibitem{Witten:1997sc}
E.~Witten, ``{Solutions of Four-Dimensional Field Theories via M-theory},''
  \href{http://dx.doi.org/10.1016/S0550-3213(97)00416-1}{{\em Nucl. Phys.}
  {\bfseries B500} (1997) 3--42},
\href{http://arxiv.org/abs/hep-th/9703166}{{\ttfamily arXiv:hep-th/9703166}}.

\bibitem{Argyres:1995jj}
P.~C. Argyres and M.~R. Douglas, ``{New Phenomena in $SU(3)$ Supersymmetric
  Gauge Theory},'' \href{http://dx.doi.org/10.1016/0550-3213(95)00281-V}{{\em
  Nucl. Phys.} {\bfseries B448} (1995) 93--126},
\href{http://arxiv.org/abs/hep-th/9505062}{{\ttfamily arXiv:hep-th/9505062}}.

\bibitem{Argyres:1995xn}
P.~C. Argyres, M.~R.~Plesser, N.~Seiberg, and E.~Witten, ``{New
  ${\mathcal{N}}\!=2$ Superconformal Field Theories in Four Dimensions},''
  \href{http://dx.doi.org/10.1016/0550-3213(95)00671-0}{{\em Nucl. Phys.}
  {\bfseries B461} (1996) 71--84},
\href{http://arxiv.org/abs/hep-th/9511154}{{\ttfamily arXiv:hep-th/9511154}}.

\bibitem{Leigh:1995ep}
R.~G. Leigh and M.~J. Strassler, ``{Exactly Marginal Operators and Duality in
  Four-Dimensional ${\mathcal{N}}\!=1$ Supersymmetric Gauge Theory},''
  \href{http://dx.doi.org/10.1016/0550-3213(95)00261-P}{{\em Nucl. Phys.}
  {\bfseries B447} (1995) 95--136},
\href{http://arxiv.org/abs/hep-th/9503121}{{\ttfamily arXiv:hep-th/9503121}}.

\bibitem{Hanany:2010qu}
A.~Hanany and N.~Mekareeya, ``{Tri-Vertices and $SU(2)$'s},''
  \href{http://dx.doi.org/10.1007/JHEP02(2011)069}{{\em JHEP} {\bfseries 02}
  (2011) 069},
\href{http://arxiv.org/abs/1012.2119}{{\ttfamily arXiv:1012.2119 [hep-th]}}.

\bibitem{Brandhuber:1995zp}
A.~Brandhuber and K.~Landsteiner, ``{On the monodromies of $\mathcal{N}=2$
  supersymmetric Yang-Mills theory with gauge group $SO(2n)$},''
  \href{http://dx.doi.org/10.1016/0370-2693(95)00986-U}{{\em Phys. Lett.}
  {\bfseries B358} (1995) 73--80},
\href{http://arxiv.org/abs/hep-th/9507008}{{\ttfamily arXiv:hep-th/9507008}}.

\bibitem{Martinec:1995by}
E.~J. Martinec and N.~P. Warner, ``{Integrable Systems and Supersymmetric Gauge
  Theory},'' \href{http://dx.doi.org/10.1016/0550-3213(95)00588-9}{{\em Nucl.
  Phys.} {\bfseries B459} (1996) 97--112},
\href{http://arxiv.org/abs/hep-th/9509161}{{\ttfamily arXiv:hep-th/9509161}}.

\bibitem{Hollowood:1997pp}
T.~J. Hollowood, ``{Strong Coupling ${\mathcal{N}}\!=2$ Gauge Theory with
  Arbitrary Gauge Group},'' {\em Adv. Theor. Math. Phys.} {\bfseries 2} (1998)
  335--355,
\href{http://arxiv.org/abs/hep-th/9710073}{{\ttfamily arXiv:hep-th/9710073}}.

\bibitem{Witten:1983tw}
E.~Witten, ``{Global Aspects of Current Algebra},''
\href{http://dx.doi.org/10.1016/0550-3213(83)90063-9}{{\em Nucl. Phys.}
  {\bfseries B223} (1983) 422--432}.

\end{thebibliography}\endgroup

\end{document}